# BloodCell-Net: A lightweight convolutional neural network for the classification of all microscopic blood cell images of the human body

[1*] Sohag Kumar Mondal; Electrical and Electronic Engineering; Khulna University of Engineering and Technology, Khulna, Bangladesh Khulna, Bangladesh; Email: ssohagkumar@gmail.com

[2*] Md. Simul Hasan Talukder; Bangladesh Atomic Energy Regulatory Authority (BAERA), E-12/A, Agargaon, Dhaka, 1207, Bangladesh; Email: simulhasantalukder@gmail.com; ORCID: [0000-0003-4592-4779]

[4] Mohammad Aljaidi; Department of Computer Science; Faculty of Information Technology; Zarqa University, Zarqa, Jordan; E-mail: mjaidi@zu.edu.jo

[5] Rejwan Bin Sulaiman; School of Computer science and Technology; Northumbria University, UK, Email: Rejwan.sulaiman@northumbria.ac.uk

[6] Md Mohiuddin Sarker Tushar; Electrical and Electronic Engineering; Bangabandhu Sheikh Mujibur Rahman Science and Technology University; mhddntshr@gmail.com

[7] Amjad A Alsuwaylimi; Department of Information Technology, College of Computing and Information Technology, Northern Border University, Arar, Saudi Arabia; E-mail: Amjad.alsuwaylimi@nbu.edu.sa

Corresponding Author:
 * 1. Md. Simul Hasan Talukder; Email: simulhasantalukder@gmail.com
   2. Sohag Kumar Mondal; Email: ssohagkumar@gmail.com

**Abstract**: Blood cell classification and counting are vital for the diagnosis of various blood-related diseases, such as anemia, leukemia, lymphoma, and thrombocytopenia. The manual process of blood cell classification and counting is time-consuming, prone to errors, and labor-intensive. Therefore, we have proposed a deep learning (DL)-based automated system for blood cell classification and counting from microscopic blood smear images. We classify a total of nine types of blood cells, including Erythrocyte, Erythroblast, Neutrophil, Basophil, Eosinophil, Lymphocyte, Monocyte, Immature Granulocytes, and Platelet. Several preprocessing steps like image resizing, rescaling, contrast enhancement and augmentation are utilized. To segment the blood cells from the entire microscopic images, we employed the U-Net model. This segmentation technique aids in extracting the region of interest (ROI) by removing complex and noisy background elements. Both pixel-level metrics such as accuracy, precision, and sensitivity, and object-level evaluation metrics like Intersection over Union (IOU) and Dice coefficient are considered to comprehensively evaluate the performance of the U-Net model. The segmentation model achieved impressive performance metrics, including 98.23% accuracy, 98.40% precision, 98.25% sensitivity, 95.97% Intersection over Union (IOU), and 97.92% Dice coefficient. Subsequently, a watershed algorithm is applied to the segmented images to separate overlapped blood cells and extract individual cells. We have proposed a BloodCell-Net approach incorporated with custom light weight convolutional neural network (LWCNN) for classifying individual blood cells into nine types. Comprehensive evaluation of the classifier's performance is conducted using metrics including accuracy, precision, recall, and F1 score. The classifier achieved an average accuracy of 97.10%, precision of 97.19%, recall of 97.01%, and F1 score of 97.10%. A 5-fold cross-validation technique is applied to split the data, which not only aids in reducing overfitting but also helps in generalizing the model.

**Keyword**: Blood Cell Classification, BloodCell-Net, U-Net, Segmentation, Light weight CNN, Watershed Algorithm

## 1. Introduction

Blood performs several vital functions in the body, including immune defense which serves as a defense mechanism against foreign elements and the transportation of oxygen, nutrients, and hormones. The blood contains cells and a portion known as plasma [29]. The blood cells comprise 45% of the total volume, while the liquid plasma constitutes the remaining 55% [30,31]. The blood cells can be classified as 3 class, red blood cell (RBC) or erythrocytes, white blood cell (WBC) or leukocytes and platelets or thrombocytes [32]. The RBCs account for 40-45% of the blood, whereas the WBCs constitute approximately 1% of the blood [33,34]. The various types of blood cells serve distinct roles within the body's organs. RBCs are primarily responsible for transporting oxygen from the lungs to tissues and organs throughout the body, facilitated by the protein hemoglobin. Meanwhile, WBCs play a critical role in the immune system, defending against infections and foreign invaders. Neutrophils, lymphocytes (including T cells and B cells), monocytes, eosinophils, and basophils comprise the diverse array of WBCs, each specializing in various aspects of immune defense and regulation [34, 35]. Basophils release histamine and heparin, which regulate allergic reactions and inflammation [37]. Eosinophils fight parasites and help regulate allergic responses. Monocytes mature into macrophages and dendritic cells, aiding in immune surveillance and presenting antigens. Lymphocytes, comprising T cells, B cells, and natural killer cells, play crucial roles in adaptive immunity [37]. Neutrophils quickly respond to bacterial infections, using phagocytosis and antimicrobial substances. A comprehensive understanding of these white blood cell types is crucial for diagnosing and treating immune-related conditions. It is observed that the proportions of neutrophils, eosinophils, lymphocytes, monocytes, and basophils in the blood are approximately 40–60%, 1–4%, 20–40%, 2–8%, and 0.5–1%, respectively [33]. Immature granulocytes, also known as band cells, are a type of WBC precursor that is produced in the bone marrow [36]. They represent an intermediate stage in the maturation of granulocytes, which include neutrophils, eosinophils, and basophils. Immature granulocytes are released into the bloodstream in response to infection or inflammation and are sometimes referred to as a "left shift" when their percentage in the blood is elevated. Platelets, also known as thrombocytes, are essential for blood clotting, which is crucial for wound healing and preventing excessive bleeding. Figure 1 illustrates the various types of blood cells. Together, these blood cells collaborate to maintain homeostasis, protect against pathogens, and ensure the proper functioning of bodily systems.

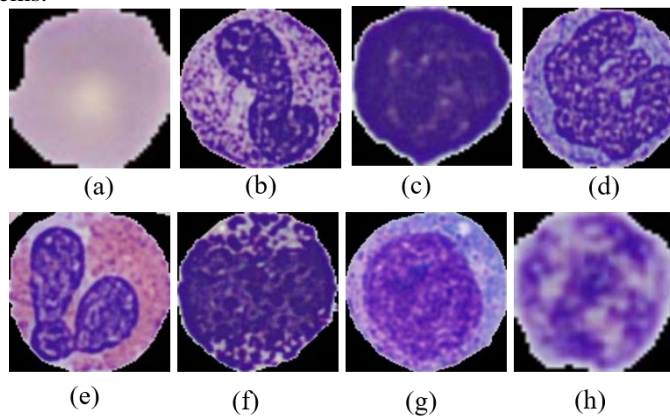

Fig. 1 Different type of blood cells, (a) Erythrocyte (Red Blood Cell), (b) Neutrophils, (c) lymphocytes, (d) monocytes, (e) eosinophils, (f) basophils (g) Immature Granulocytes and (h) Thrombocytes (Platelets).

The peripheral blood smear is a standard laboratory examination that offers physicians extensive insights into a patient's overall health status [17]. Leukemia and malaria are diseases characterized by alterations in WBC count, emphasizing the significance of early diagnosis [28]. Additionally, Patients can be evaluated for various health conditions, including immune system disorders and the presence of cancerous cells [38]. It offers both qualitative and quantitative evaluations of blood constituents, primarily focusing on cells and platelets. The manual blood cell count relies on the microscopic examination of the blood smear by the analyst, who distinguishes between subtypes primarily based on the morphological features including size, shape, texture, nucleus of the cell nucleus and cytoplasm [40]. Nonetheless, this process can be time-consuming and prone to errors if the microscopists are not adequately trained [39]. Furthermore, as this hematological assessment is a routine test, it frequently experiences high demand in clinical laboratories, leading to increased workload and impacting performance [17]. Thus, the implementation of a computer-aided diagnosis (CAD) system is necessary to provide diagnostic assistance in the laboratory.

The CAD system for blood cell classification can be decomposed as four parts, segmentation, ROI extraction, Feature Extraction and classification. Typically, cell segmentation poses a challenge in tissue samples. However, this process is simpler in cell smears due to the distinct appearance of the dark nucleus in staining blood smear. Performing segmentation before classification in blood cell classification tasks offers several advantages. It enhances accuracy through the isolation of individual cells and extraction of specific features, thereby improving classification and counting capabilities. This approach reduces computational complexity and enhances robustness by eliminating backgrounds and non-essential substances. Overall, segmentation before classification enhances the accuracy, efficiency, and robustness of blood cell classification tasks. In traditional image segmentation techniques, including manual thresholding [47], OTSU binarization [48], fuzzy c-means (FCM) [49], active contours [50], and watershed algorithm [51], are employed. Following the advent of deep learning (DL), numerous segmentation techniques have been developed, including the fully connected network (FCN) [52], U-Net [53], Faster-RCNN [54], YOLO [55], among others. These approaches have demonstrated superior performance compared to traditional methods for blood cell segmentation.

In conventional image processing techniques, the images are classified based on different features like histogram of gradients (HOG), colors, texture, geometric, edges, and statistical features [46]. Due to the benefits of artificial intelligence in image analysis, various machine learning (ML) techniques have been examined for the classification and segmentation of leukocytes. These approaches span from conventional methods like support vector machines (SVM) [42] and Naïve Bayesian [43] to advanced algorithms such as DL models [44,45]. Transfer learning has introduced an additional dimension to the classification of blood cells by accelerating the training speed. Researchers utilized different transfer learning models like AlexNet [56], ResNet [15], VGG [44], GoogLeNet [57] for automated blood cell classification.

This study presents a novel BloodCell-Net scheme for the segmentation and classification of nine types of blood cells, including Erythrocyte, Erythroblast, Neutrophil, Basophil, Eosinophil, Lymphocyte, Monocyte, Immature Granulocytes, and Platelet. We utilized various data preprocessing techniques, including image resizing, normalization, histogram equalization and augmentation. These preprocessing steps enhance model performance during data preparation. Image resizing ensures uniformity in image dimensions, aiding efficient processing. Normalization of pixel values stabilizes training and improves the model's learning ability. Histogram equalization enhances feature visibility for better extraction. Augmentation techniques, such as rotation and scaling, increase dataset diversity, reducing overfitting. Together, these steps optimize the dataset for robust model training. To segment the blood cells from the background, we employed the U-Net model, specifically designed for segmenting medical images. The pixel-wise segmentation model enhances classifier performance by isolating the non-important background from the region of interest (ROI). In microscopic thin blood smear images, the overlapping of blood cells is a common occurrence, presenting challenges in their separation. Hence, the Watershed algorithm is utilized to effectively separate overlapping cells and extract the ROIs from the segmented images. Here in our study the ROIs are the single blood cells. Finally, we proposed a custom sequential LWCNN architecture to classify the blood cells. The contribution of this research work is given below.

- Preparation of pre-processed dataset having nine types of blood smear dataset.
- Introducing data scaling, histogram equalization & augmentation.
- Implementing a precise segmentation approach to accurately delineate the regions of interest within the images.
- Application of the watershed algorithm to separate overlapping cells and extract the ROIs from segmented images.
- Proposing an effective LWCNN architecture for performing classification tasks.
- Presenting a novel BoodCell-Net approach to detecting nine types of blood cells.
- Evaluation of the model with well-known performance metrics.

The proposed paper is structured as follows: Section 1 presents the introduction, while Section 2 provides a review of the literature. The proposed framework model is detailed in Section 3, with the results and discussion presented in Section 4. Finally, Section 5 offers the conclusion.

## 2. Literature Review

Artificial Intelligence is an emerging technology that is being used widely in agriculture [93], medical image processing [94] and healthcare [95]. Following the research community extensively employs both traditional machine learning (TML) and deep learning (DL) models for the automated classification of blood cells [8, 17, 64, 65]. Many researchers divide this problem into two parts, segmentation and classification, in order to find the appropriate solution [9,13]. Segmentation enhances classifier performance by removing non-relevant parts from the images. Segmentation techniques, such as thresholding, morphological operations, and machine learning, are commonly employed in the research cited as [69, 70, 71, 67, 66]. Nee et al. introduced a segmentation technique for WBCs in acute leukemia bone marrow images, leveraging thresholding and morphological operations [71]. The research article cited as [70] implemented a segmentation process for blood cells utilizing thresholding and Canny edge detection, followed by enhancing local and global details of the output through morphological operations. The authors claimed an average segmentation accuracy of 87.9% in the public human RBC dataset. Both segmentation and classification of WBCs are performed by Pešić to detect Acute Lymphoblastic Leukemia. The author employed the Otsu thresholding technique twice on the H channel in the HSI color space. In machine learning for blood cell segmentation both supervised and unsupervised techniques are used [66, 68, 67, 72]. Tran et al. introduced a CNN based model called SegNet for the classification of blood cells. Notably, this model was initialized with weights derived from the VGG-16 network. Among the three types of blood cells, the researchers considered RBCs and WBCs, while omitting platelets from their analysis. U-Net, a specialized CNN-based model designed specifically for medical image segmentation, was utilized by Zhang et al. in their article [68]. The proposed deformable U-Net (dU-Net) demonstrated superior performance on both binary segmentation and multiclass semantic segmentation tasks [68]. Beside supervised ML models, the unsupervised ML models are also used among researcher communities for the blood cell segmentation [67, 72]. The research cited as [67, 72] is utilized and discusses k-means clustering image segmentation technique to segment the WBCs in their paper. Zheng et. al. in [72], proposed a novel technique for separating the overlapping cells after segmentation based on the touching-cell clump splitting technique.

Classification plays a pivotal role in comprehending the blood profile and estimating cell density by counting the number of cells. Some authors have employed traditional machine learning techniques, while others have utilized deep learning approaches such as custom CNNs, transfer learning, and hybrid models for blood cell classification [17, 3, 19, 28]. Hegde et al. introduced a leukemia detection system employing an SVM classifier [42]. The authors incorporated various features such as shape, color, and texture into their analysis. The proposed approach yields encouraging outcomes in differentiating between normal and abnormal WBCs, potentially reducing the workload for pathologists significantly. An automatic leukocytes classification system is built by extracting morphological features based on the Naïve Bayes Classifier by Gautam et. al. [73]. The features which are extracted by the authors are area, eccentricity, perimeter and circularity of leukocyte nucleus. This research article claimed an accuracy of 80.88% for this particular task. On the contrary, Prinyakupt et al. employed and evaluated the performance of two distinct models, namely the linear classifier and Naïve Bayes, for the WBC classification system [43]. They show that the linear classifier outperforms the Naïve Bayes.

In addition to traditional machine learning techniques, researchers have also incorporated deep learning models into the task of blood cell classification [3,24,5,11,18]. A new 33-layer CNN model called WBCNet model that can extract features and classify WBCs from microscopic WBC images is proposed by Jiang et. al [3]. This research integrated the batch normalization algorithm, residual convolution architecture, and enhanced activation functions to enhance the classification score. The study referenced as [11] proposes a CNN-based approach utilizing Genetic Algorithm (GA) to classify four types of leukocytes, achieving 99% training accuracy and 91% validation accuracy. Su et. al. highlighted and extracted three types of features called geometrical features, color features, and LDP-based texture features for WBC classification [14]. The extracted features are fed into three different kinds of neural networks to recognize the types of the WBCs. The simulation results demonstrated that the proposed system for classifying WBCs is highly competitive when compared to several existing systems. A Regional Based CNN (R-CNN) is proposed by Kutlu et. al. for classifying WBCs [18]. The proposed model also compared with four transfer learning models called AlexNet, VGG16, GoogLeNet, ResNet50, and claimed the proposed R-CNN outperformed the state-of-the-arts. Feature fusion is the process of combining information from multiple sources or representations into a single representation to improve performance [74]. A novel feature fusion based deep learning framework for white blood cell classification is proposed by Dong et. al. [22]. The authors fused deep learning features with artificial features to generate the final features for WBC classification.

Various transfer learning models such as AlexNet, MobileNetV2, DenseNet161, and DarkNet-53 are employed for the classification of blood cells and leukemia [75,76,77,80,81]. Loey et al. utilized a pre-trained deep CNN, namely AlexNet, to extract features before classifying WBCs as either healthy or infected by leukemia [75]. Tamang et al. assessed various state-of-the-art models using multiple evaluation metrics such as accuracy, precision, recall, F1 score and training time ultimately recommending DenseNet161 for blood cell classification [76]. Furthermore, advanced optimization methods like normalization, mixed-up augmentation, and label smoothing were applied to DenseNet to further enhance its performance. In another study, Yang et al. trained three distinct CNN architectures by initializing the pretrained parameters on ImageNet and recommended MobileNetV2 for blood cell classification [81]. This research claimed 89.40% accuracy and 91.60% precision on the BCCD dataset. The study conducted by Saleem et al. employed a generative adversarial network (GAN) to augment image data for subsequent classification using DarkNet-53 [80]. Less important features are dropped, and the more important features are selected using principal component analysis (PCA) technique. Subsequently, these features are fused using various machine learning models. Oumaima Saidani et al. [88] introduced one optimized CNN and four transfer learning models, namely MobileNetV2, VGG16, InceptionV3, and ResNet50 to classify five types of human blood cells collected from the IEEE data port. The study also integrated color conversion and augmentation preprocessing. The optimized CNN model achieved 99.86% accuracy.

A few studies were conducted on eight types of blood cells. Recently, Karnika Dwivedi et al. [86] proposed a CNN-based architecture called MicrocellNet to classify eight types of blood cells. Their approach achieved 98.76% validation accuracy and 97.65% test accuracy. Similarly, another most recent work was carried out on eight types of blood cell classification by the researcher named Hüseyin Fırat et al. [87]. The author presented a hybrid method combination of the Inception module, pyramid pooling module (PPM), and depth wise squeeze-and-excitation block (DSEB) to classify blood cells. The study used three datasets named four classes (BCCD dataset), five classes (Raabin WBC dataset), and eight classes and achieved 99.96%, 99.22%, and 99.72% accuracy, respectively. But they did not use K-fold validation to check the robustness of the method.

Table 1 presents a comparative summary of literature reviews. In the literature survey, three types of public datasets have been found that are used in different studies to classify blood cells in the human body. The dataset consists of four classes (BCCD), five classes (WBC), and eight classes. Most of the research focused on four and five classes of blood classification. The majority of the research concentrated on transfer learning for classifying blood cells, while a small study used statistical features and a normal CNN model. But notably, transfer learning exhibited impressive accuracy in WBC and BCCD data classification. Many studies did not apply cross-validation to check for robustness. It is important to highlight that only two studies were performed on eight classes of blood cell classification. The eight-class dataset excluded erythrocyte types of cells. Additionally, no study has included all types of human blood cells, including erythrocytes. Fine tuning of histogram equalization, segmentation, augmentation, watershed algorithm, and light weight CNN were also not employed in any previous work. In our study, we addressed all the points by preparing nine types of datasets and adding advanced preprocessing techniques, as well as a light-weight CNN with suppressive performance.

Table 1. Comparative literature review summary.

| Ref. | Classifier | Dataset | Accuracy |
|---|---|---|---|
| [90] | AlexNet, VGG Net 13, 11, ResNet 18, 34, 50, SqueezeNet 10, 11, DenseNet 121, 161. | BCCD dataset (four classes) | 100% (DenseNet 161) |
| [92] | DT, RF, SVM, k-NN, CNN, VGG16, ResNet15 | IEEE data port (3539images) Five classes | 97% (RF) |
| [91] | CNN, ELM, AlexNet, GoogleNet, VGG-16, and ResNet | BCCD dataset (four classes) | 96.03% |
| [89] | CNN | Kaggle | 98.55% |
| [88] | Optimized CNN model | IEEE data port (3539images) (Five classes) | 99.86% |
| [86] | CNN based MicrocellNet | Mendeley dataset (Eight classes) | 97.65% |
| [87] | Hybrid method combination of the Inception module, pyramid pooling module (PPM), and depth wise squeeze-and-excitation block | 1. Mendeley dataset (Eight classes) 2. IEEE data port (3539images) (Five classes) 3. BCCD dataset (four classes) | 1. 99.72% 2. 99.22%, 3. 99.96% |

## 3. Materials and Methods

In this section, we introduce the proposed BloodCell-Net framework for the classification of blood cells into nine distinct classes, as depicted in Figure 2. The framework comprises several fundamental components, namely data preprocessing, segmentation, ROI extraction, and classification. Each of these components is elaborated upon sequentially as follows.

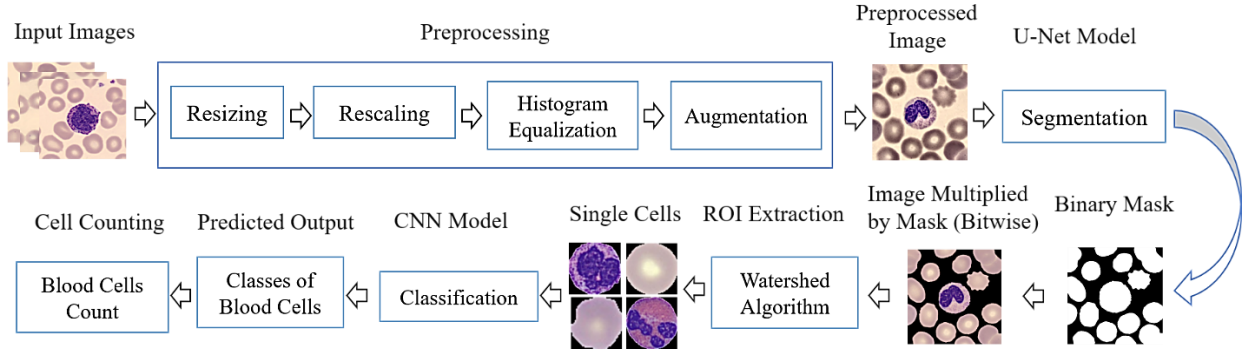

Fig. 2. Block diagram of the proposed BloodCell-Net framework for blood cell classification.

### 3.1. Dataset Description

The microscopic peripheral blood cell images dataset has been collected from a trustworthy Mendeley repository [85]. The dataset consists of 17,092 images of distinct cells of normal function which were obtained in the Hospital Clinic of Barcelona's Core Laboratory using the analyzer CellaVision DM96. It is categorized into eight groups: neutrophils, eosinophils, basophils, lymphocytes, monocytes, immature granulocytes (promyelocytes, myelocytes, and metamyelocytes), erythroblasts and platelets or thrombocytes. Expert clinical pathologists annotated the 360 x 363-pixel JPG photos. The blood samples were taken from people who were clear of infections, hematologic or oncologic diseases, and pharmaceutical treatments at the time of the photo session.

In this study, we extracted erythrocyte cells from the whole dataset using a watershed approach, which we named after the erythrocyte class. So, the dataset of our study consists of nine types microscopic peripheral blood cells namely, Erythrocyte, Erythroblast, Neutrophil, Basophil, Eosinophil, Lymphocyte, Monocyte, Immature Granulocytes, and Platelet. The distribution and visual appearance of the dataset is shown in Table 2 and Figure 3 respectively.

Table 2. Distribution of images in each class.

| Name of the Class | Number of the images |
|---|---|
| Erythrocyte | 1200 |
| Erythroblast | 1168 |
| Neutrophil | 1133 |
| Basophil | 969 |
| Eosinophil | 1186 |
| Lymphocyte | 1131 |
| Monocyte | 999 |
| Immature Granulocytes | 1134 |
| Platelet | 1204 |

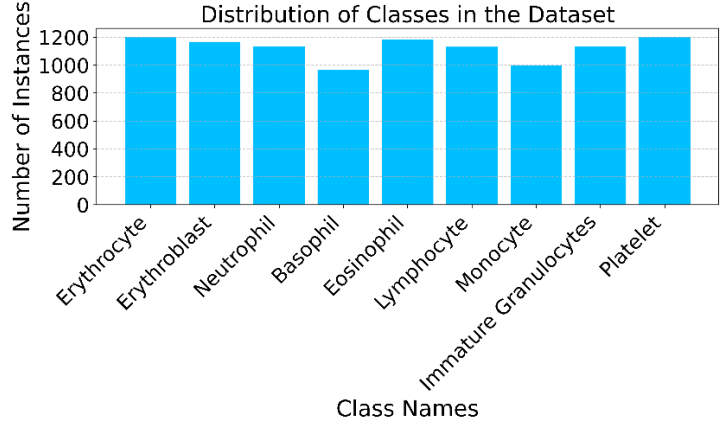

Figure 3. Bar plot for Dataset distribution.

### 3.2. Preprocessing
Image preprocessing is a crucial step to enhance the performance of subsequent models. In this step, we utilize image resizing, rescaling, histogram equalization, and augmentation.

### 3.2.1. Image Resizing
In the dataset, we encountered images with varying sizes, which are not conducive to further processing. Therefore, we resize the images to 224×224 px to standardize the dimensions for consistent and efficient processing. Linear interpolation techniques have been employed in this study. The mathematical expression of image resizing by linear interpolation is given by the following equation-

$$I_{resized}(x', y') = (1 - \alpha) \cdot I_{orginal}(x_1, y') + \alpha \cdot I_{orginal}(x_2, y') \qquad (1)$$

Where, $\alpha$= fractional part of $x'$ and $x_1, y_1$ are the neighboring pixel indices. $I_{resized}(x', y')$ is the pixel intensity of the resized image and $I_{orginal}(x, y')$ is the pixel intensity of original image.

### 3.2.2. Rescaling
Rescaling or normalization in image data preprocessing is crucial for ensuring that pixel values across images are on a similar scale. This process is essential for machine learning models, particularly deep neural networks, as it prevents certain features or channels from dominating others during training. Normalization also helps in stabilizing and accelerating the convergence of optimization algorithms, such as gradient descent, by ensuring that gradients are not disproportionately influenced by large pixel values. Furthermore, normalization aids in improving the interpretability of CNN architectures by facilitating more consistent feature representations across layers, thus enabling better generalization and performance on image classification, segmentation, and other computer vision tasks. As the images in our dataset are RGB images, the pixel values range from 0 to 255. Therefore, we divided the pixel values by 255 to normalize them and bring all pixels within the range of 0 to 1. The expression normalization can be expressed by the following equation-

$$x' = \frac{x - x_{min}}{x_{max} - x} \qquad (2)$$

$x$ is the original value and $x'$ is normalized value.

### 3.2.3. Histogram Equalization
The images in the dataset are captured under various light conditions, leading to variations in brightness and contrast. That is why an image enhancement technique is required to equalize the histogram of the images. We utilized Contrast Limited Adaptive Histogram Equalization (CLAHE) technique enhances image contrast by dividing the image into smaller regions, known as tiles, and applying histogram equalization to each tile independently [83]. This adaptive approach ensures that contrast enhancement is tailored to the local characteristics of the image, making it effective for images with non-uniform illumination or regions of varying contrast. To prevent over-amplification of noise in low-contrast areas, CLAHE incorporates a contrast limiting mechanism that clips the histogram beyond a specified threshold. Additionally, bilinear interpolation is used to blend adjacent tiles, smoothing out any discontinuities and artifacts.

### 3.2.4. Augmentation

Augmentation is a technique to increase the number of training data from the existing ones using different morphological operations. In this study, seven augmentation strategies were applied that are rotation, width shift, height shift, shear, zoom, flip, and fill.

Rotation: To reduce the model's sensitivity to object positioning, random adjustments were made to the image angles, which ranged from -20 to 20 degrees. This tactic facilitates the model's ability to identify features across a range of orientations. A 2D affine transformation can be used to depict the rotation of an image.

$$\begin{bmatrix} x' \\ y' \end{bmatrix} = \begin{bmatrix} \cos(\theta) & -\sin(\theta) \\ \sin(\theta) & \cos(\theta) \end{bmatrix} \begin{bmatrix} x \\ y \end{bmatrix} \quad (3)$$

Where, $\theta$ is the angle of rotation $(x, y)$ are the initial coordinates in the image and $(x', y')$ are the coordinates following rotation.

**Width and Height Shift:** Both height shift and width shift entail translating the image both vertically and horizontally. These modifications assist the model in adjusting to shifts in the locations of items within the frame.
These transformations have the following mathematical expression:

$$\begin{bmatrix} x' \\ y' \end{bmatrix} = \begin{bmatrix} x \\ y \end{bmatrix} + \begin{bmatrix} \Delta x \\ \Delta y \end{bmatrix} \quad (4)$$

In this expression, $(x, y)$ are the main coordinates in the image, $(x', y')$ are the coordinates after employing the width and height shift transformations, and $\Delta x, \Delta y$ are the shift values for width and height, respectively. Adjusts from -0.05 to 0.05, adds randomness, normalizes based on the size of the input image, and facilitates the model's ability to adjust to new data.

**Shear:** The model's flexibility is increased in response to minute modifications that occur in real-world data by applying shear transformations between 0 and 0.05. Shearing is shown as follows:

$$\begin{bmatrix} x' \\ y' \end{bmatrix} = \begin{bmatrix} 1 & \lambda \\ 0 & 1 \end{bmatrix} \begin{bmatrix} x \\ y \end{bmatrix} \quad (5)$$

In this equation, (x, y) are the initial coordinates in the image, $(x', y')$ are the coordinates following the shear transformation, and $\lambda$ is the shearing factor, ranging from 0 to 0.05.

**Zoom:** To allow the model to adapt to variations in object size, zoom transformations between 0 and 0.05 have been added. This enhances the model's perception of features at varying magnifications. The act of zooming is defined as:

$$\begin{bmatrix} x' \\ y' \end{bmatrix} = \begin{bmatrix} \alpha & 0 \\ 0 & \beta \end{bmatrix} \begin{bmatrix} x \\ y \end{bmatrix} \quad (6)$$

where $(x, y)$ are the initial coordinates in the image, $(x', y')$ are the coordinates following employing the zoom transformation, and $(\alpha, \beta)$ are the scaling factors for width and height, respectively, which vary from 1 to 1.2 to indicate a zoom range of 0 to 0.05.

**Horizontal Flip:** To increase the model's comprehension of symmetry, a horizontal flip augmentation is used. As a result, the model can identify features both in their original and reversed states. A straightforward process called "horizontal flipping" flips the values of the pixels along the vertical axis.

$$I_{flipped}(x, y) = I(-x, y) \quad (7)$$

The pixel value at location $(x, y)$ in the flipped image is represented by the equation $I_{flipped}(x, y)$ and the pixel value at the mirrored position in the original image is indicated by $I(-x, y)$.

**Fill Mode:** "Fill Mode" is a technique for managing portions of a picture that could appear after geometric operations like rotation, shifting, or shearing. After an image has been converted, if some pixels are not filled with their original values, there can be blank or undefinable parts. For a pixel at coordinates $(x', y')$ in the transformed image, the corresponding pixel value $I(x', y')$ can be approximated using nearest-neighbor interpolation as

$$(I'((x', y') = I(|(x'|, |y'|) \quad (8)$$

## 3.3. Image Segmentation

In this study, a U-Net model is deployed to segment the peripheral blood cell images. The U-Net architecture leverages encoder-decoder skip connections and multi-resolution feature fusion to achieve accurate segmentation of microscopic blood cell images. By preserving spatial details and capturing hierarchical features, the model demonstrates robust performance in segmenting complex cell structures, contributing to advancements in medical image analysis. The encoder portion of the network consists of four convolutional blocks followed by max-pooling layers to down-sample the spatial dimensions of the input image while increasing the number of feature maps. Each convolutional block applies two 3×3 convolutional layers with ReLU activation, preserving spatial information and extracting increasingly abstract features. The number of filters doubles with each block, starting from 64 and ending with 512. Following the encoder, a middle convolutional block further processes the feature maps, aiming to capture high-level semantic information. This block has the same structure as the previous ones but operates on the highest resolution feature maps extracted by the encoder. The decoder portion of the network mirrors the encoder's structure, but with up-sampling operations to gradually restore the spatial dimensions of the feature maps. After each up-sampling operation, concatenation is performed between the up-sampled feature maps and the corresponding feature maps from the encoder, enabling the network to recover spatial details lost during down-sampling. Each concatenated feature map undergoes convolutional operations to refine the segmentation output. The number of filters decreases with each block, mirroring the pattern of the encoder. The final layer of the network consists of a 1×1 convolutional layer with sigmoid activation, producing the segmentation mask for the input image. The sigmoid activation function ensures that the output values are in the range [0, 1], representing the probability of each pixel belonging to the foreground class (blood cell). The model is trained using the binary cross-entropy loss function and the Adam optimizer. Binary cross-entropy is suitable for binary classification tasks like image segmentation, where each pixel is classified as either foreground (blood cell) or background. In Fig. 4 the structure of U-Net architecture is depicted, where table 3 is constructed with the architecture of U-Net model.

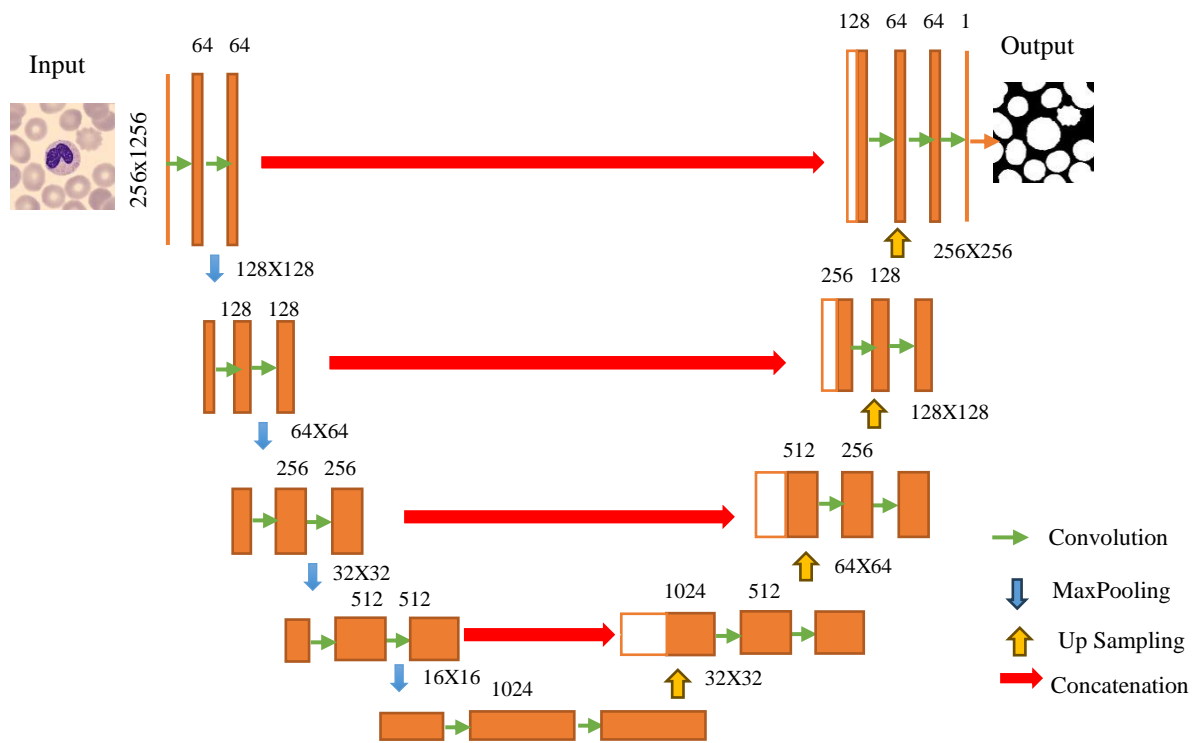

Fig. 4. Architecture of U-Net model.

Table 3. Summary table of U-Net architecture.

| Layer (Type) | Feature Map | No. of Parameters | Kernel Size | Activation Function |
|---|---|---|---|---|
| InputLayer | (None, 256, 256, 3) | 0 | - | - |
| Conv2D_1 | (None, 256, 256, 64) | 1792 | (3, 3) | ReLU |
| Conv2D_2 | (None, 256, 256, 64) | 36928 | (3, 3) | ReLU |
| MaxPooling2D_1 | (None, 128, 128, 64) | 0 | (2, 2) | - |
| Conv2D_3 | (None, 128, 128, 128) | 73856 | (3, 3) | ReLU |
| Conv2D_4 | (None, 128, 128, 128) | 147584 | (3, 3) | ReLU |
| MaxPooling2D_2 | (None, 64, 64, 128) | 0 | (2, 2) | - |
| Conv2D_5 | (None, 64, 64, 256) | 295168 | (3, 3) | ReLU |
| Conv2D_6 | (None, 64, 64, 256) | 590080 | (3, 3) | ReLU |
| MaxPooling2D_3 | (None, 32, 32, 256) | 0 | (2, 2) | - |
| Conv2D_7 | (None, 32, 32, 512) | 1180160 | (3, 3) | ReLU |
| Conv2D_8 | (None, 32, 32, 512) | 2359808 | (3, 3) | ReLU |
| MaxPooling2D_4 | (None, 16, 16, 512) | 0 | (2, 2) | - |
| Conv2D_9 | (None, 16, 16, 1024) | 4719616 | (3, 3) | ReLU |
| Conv2D_10 | (None, 16, 16, 1024) | 9438208 | (3, 3) | ReLU |
| UpSampling2D_1 | (None, 32, 32, 1024) | 0 | (2, 2) | - |
| Concatenate_1 | (None, 32, 32, 1536) | 0 | - | - |
| Conv2D_11 | (None, 32, 32, 512) | 7078400 | (3, 3) | ReLU |
| Conv2D_12 | (None, 32, 32, 512) | 2359808 | (3, 3) | ReLU |
| UpSampling2D_2 | (None, 64, 64, 512) | 0 | (2, 2) | - |
| Concatenate_2 | (None, 64, 64, 768) | 0 | | |
| Conv2D_13 | (None, 64, 64, 256) | 1769728 | (3, 3) | ReLU |
| Conv2D_14 | (None, 64, 64, 256) | 590080 | (3, 3) | ReLU |
| UpSampling2D_3 | (None, 128, 128, 256) | 0 | (2, 2) | - |
| Concatenate_3 | (None, 128, 128, 384) | 0 | - | - |
| Conv2D_15 | (None, 128, 128, 128) | 442496 | (3, 3) | ReLU |
| Conv2D_16 | (None, 128, 128, 128) | 147584 | (3, 3) | ReLU |
| UpSampling2D_4 | (None, 256, 256, 128) | 0 | (2, 2) | - |
| Concatenate_4 | (None, 256, 256, 192) | 0 | - | - |
| Conv2D_17 | (None, 256, 256, 64) | 110656 | (3, 3) | ReLU |
| Conv2D_18 | (None, 256, 256, 64) | 36928 | (3, 3) | ReLU |
| Conv2D_19 | (None, 256, 256, 1) | 65 | (1, 1) | Sigmoid |

## 3.4. Watershed Algorithm

In this study, we employed the watershed algorithm to detect and isolate individual blood cells from microscopic images. By utilizing morphological operations and distance transformation, the watershed algorithm delineated cell boundaries accurately. Subsequently, connected component analysis was performed to identify individual cells, and those with an area greater than a predefined threshold were saved as separate images for further analysis. This method ensured precise cell segmentation, vital for subsequent blood cell classification tasks. In Figure 5, the image depicts the state before and after segmentation by the watershed algorithm. The working way of watershed algorithm is shown in algorithm 1.

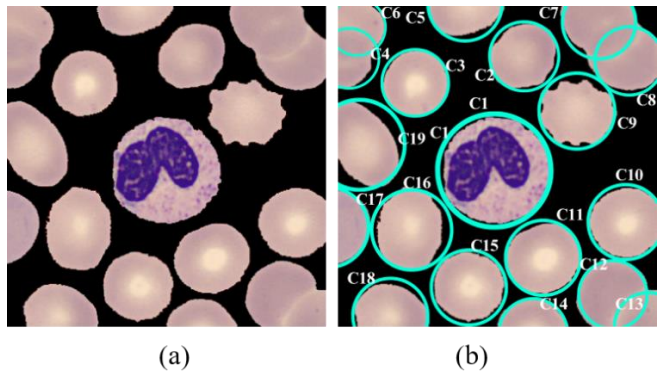

Fig 5 (a) Image before segmented by watershed algorithm, (b) after segmented by watershed algorithm.

| | **Algorithm 1. Watershed Algorithm** |
|---|---|
| 1 | **Input**: Load the image |
| 2 | **Output**: Segmented Image |
| 3 | Compute the gradient magnitude of the image |
| 4 | Find the marker |
| 5 | Initialize a priority queue to hold pixels sorted by gradient magnitude |
| 6 | Initialize labels matrix with all pixels labeled as unprocessed (-1) |
| 7 | Initialize a priority queue (heap) of basins |
| 8 | **For** each pixel in the image: |
| 9 |     Add pixel to the priority queue |
| 10 | **End** |
| 11 | **While** priority queue =!0 **do** |
| 12 |     Pop pixel with smallest gradient magnitude from the queue |
| 13 |     **If** pixel == marker |
| 14 |         Add pixel to the basin priority queue |
| 15 |     **else** |
| 16 |         Label pixel with marker's label |
| 17 |         **For each** neighboring pixel: |
| 18 |             **If** neighboring pixel == unlabeled |
| 19 |                 Label neighboring pixel with marker's label |
| 20 |                 Add neighboring pixel to the priority queue |
| 21 |             **Else if** neighboring pixel == within another different marker |
| 22 |                 Merge basins associated with these markers |
| 23 |                 Add the merged basin to the basin priority queue |
| 24 |             **End** |
| 25 |         **End** |
| 26 |     **End** |
| 27 | **End** |
| 28 | Generate segmented image based on labeled pixels |

### 3.5. Data Splitting

K-fold cross-validation is a technique used in machine learning to assess the performance of predictive models by partitioning the dataset into k equal-sized subsets. The model is trained on k-1 folds and evaluated on the remaining fold, with this process repeated k times. By averaging performance metrics across the folds, k-fold cross-validation provides an unbiased estimate of the model's performance, reducing variance and enhancing robustness. It allows for effective parameter tuning, maximizes data utilization, and provides a more comprehensive assessment of the model's generalization ability. Overall, k-fold cross-validation is a valuable tool for model evaluation, improving reliability, and ensuring optimal performance in machine learning tasks. In our study into blood cell classification utilizing the Mendeley repository dataset, we implemented a 5-fold cross-validation technique. This entailed partitioning the dataset into five subsets, with four utilized for model training and the remaining one for testing in each iteration. This approach facilitated a thorough assessment of the model's performance, maintaining a balanced utilization of the dataset. Fig 6 illustrates the data splitting mechanism employed for 5-fold cross-validation.

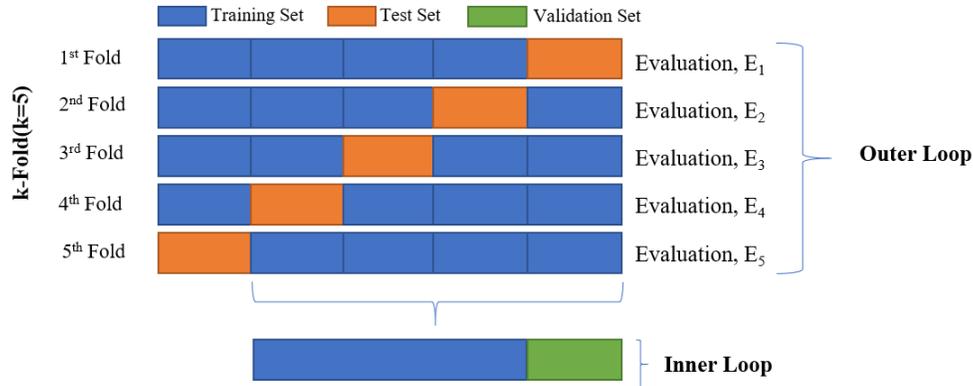

Fig 6. Data splitting technique of 5-fold cross validation technique.

### 3.6. Proposed LWCNN model

The proposed LWCNN model for blood cell classification exhibits a sophisticated architecture designed to effectively learn and distinguish features from input images. The model begins with a series of convolutional layers, each followed by max-pooling layers, which enable the extraction of hierarchical features while reducing spatial dimensions. Utilizing "ReLU" activation functions, these convolutional layers introduce non-linearity to the model, enhancing its capability to capture complex patterns. To prevent overfitting, dropout layers are strategically incorporated after each max-pooling operation, randomly dropping a fraction of the neuron units during training. This regularization technique helps to improve the model's generalization performance by reducing the likelihood of learning noise or irrelevant features. The final layers of the model consist of densely connected layers, culminating in a "Softmax" activation function that produces probability distributions over the output classes. With a total of 653,129 trainable parameters, our LWCNN model demonstrates a considerable capacity to learn discriminative representations from the input blood cell images, paving the way for accurate classification results. Fig 7 shows the proposed LWCNN architecture where the Table 4 provides a detailed overview of its structural components.

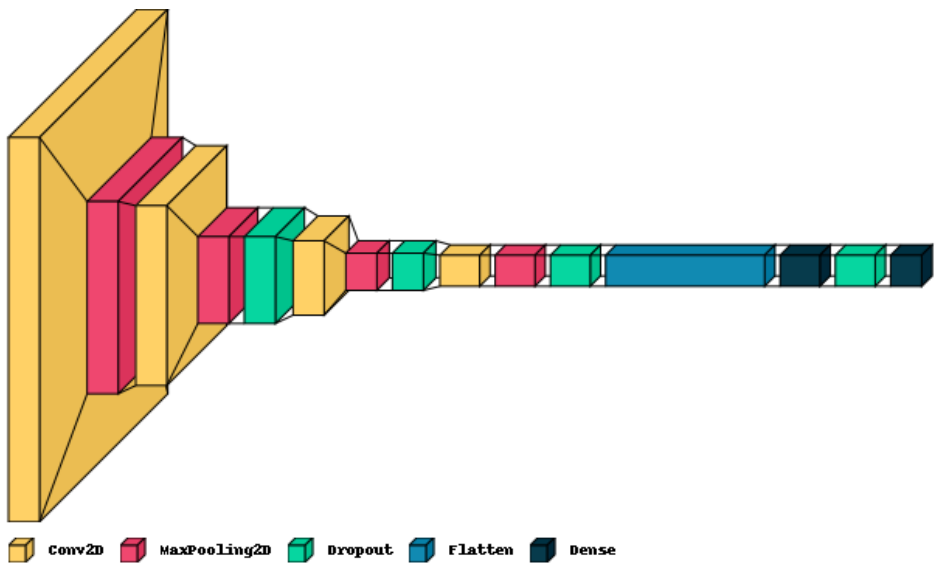

Fig 7. The proposed LWCNN architecture for blood cell classification.

Table 4. Summary table of the proposed CNN architecture.

| Layer (Type) | Feature Map | No. of Parameters | Kernel Size | Activation Function |
|---|---|---|---|---|
| Conv2D | (None, 62, 62, 32) | 896 | (3, 3) | ReLU |
| MaxPooling2D | (None, 31, 31, 32) | 0 | (2, 2) | - |
| Conv2D | (None, 29, 29, 64) | 18,496 | (3, 3) | ReLU |
| MaxPooling2D | (None, 14, 14, 64) | 0 | (2, 2) | - |
| Dropout | (None, 14, 14, 64) | 0 | - | - |
| Conv2D | (None, 12, 12, 128) | 73,856 | (3, 3) | ReLU |
| MaxPooling2D | (None, 6, 6, 128) | 0 | (2, 2) | - |
| Dropout | (None, 6, 6, 128) | 0 | - | - |
| Conv2D | (None, 4, 4, 256) | 295,168 | (3, 3) | ReLU |
| MaxPooling2D | (None, 2, 2, 256) | 0 | (2, 2) | - |
| Dropout | (None, 2, 2, 256) | 0 | - | - |
| Flatten | (None, 1024) | 0 | - | - |
| Dense | (None, 256) | 262,400 | - | ReLU |
| Dropout | (None, 256) | 0 | - | - |
| Dense | (None, 9) | 2,313 | - | Softmax |

## 4. Result and Discussion

CLAHE is applied to enhance the contrast of our dataset images, effectively highlighting the distinguishing features between subjects and background. This enhancement plays a pivotal role in enabling the U-Net model to accurately segment blood cells amidst noisy backgrounds. A randomly selected image sample from our dataset is displayed in Fig 8 alongside its histogram-equalized version. The RGB channel distribution is also depicted, illustrating that the image's RGB channel distribution exhibits sharp peaks prior to histogram equalization, while the channels become more uniform after applying CLAHE.

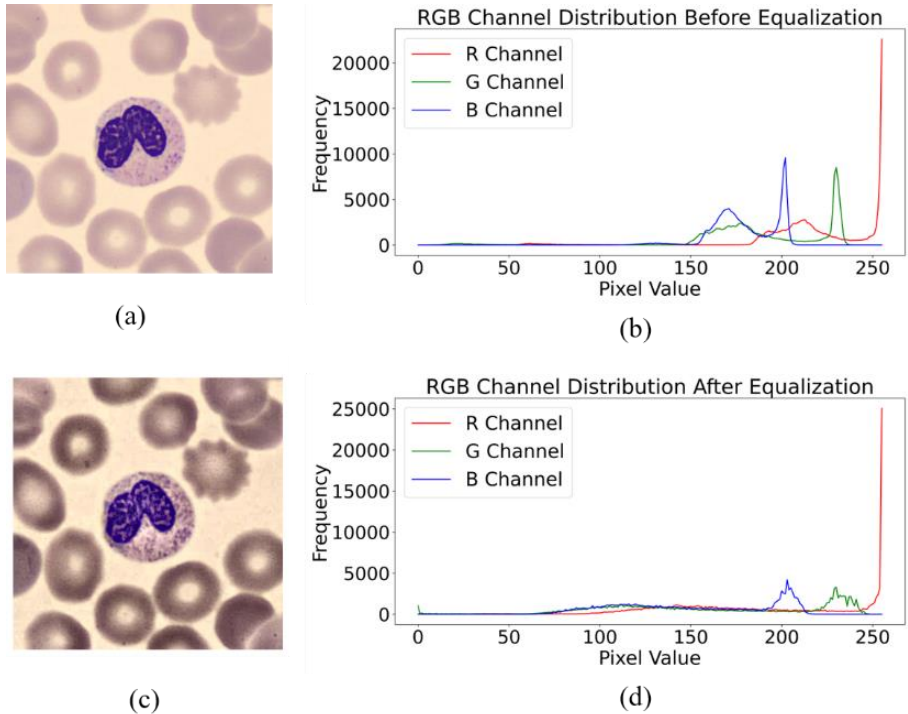

Fig 8. (a) Original image, (b) RGB channel distribution before applying CLAHE, (c) Histogram equalized image, (d) RGB channel distribution after applying CLAHE.

Several evaluation metrics are used, each focusing on a different component of our proposed model's performance, thereby assessing how well it classifies the blood cell.

**Accuracy**: Accuracy indicates how well the model's predictions correspond with the actual outcomes. It basically evaluates the overall performance of the model. It is computed with the following formula:
$$Accuracy = \frac{TP+TN}{TP+TN+FP+FN} \quad (9)$$

**Precision:** This assesses how well the model can lower false positives, demonstrating the accuracy with which types of blood cell cases are confirmed. It is described as:
$$Precision = \frac{TP}{TP+FP} \quad (10)$$

**Recall:** Recall presents the capacity of the model to detect every instance of a real blood cell and is as follows:
$$Recall = \frac{TP}{TP+FN} \quad (11)$$

**F1 Score:** The trade-offs between recall and precision are balanced by the F1-Score. Calculated as the harmonic mean of these two measures, it equals-
$$F1 = 2 \times \frac{Precision \times Recall}{Precision+Recall} \quad (12)$$

Where, TP= True Positive; TN= True Negative; FP=False Positive; FN= False Negative

Pixel-level and object-level metrics are two types of evaluation metrics used for assessing segmentation models like U-Net. Pixel accuracy, precision, and recall (or sensitivity) are examples of pixel-level metrics, while Intersection over Union (IoU) and Dice coefficient are examples of object-level metrics. Pixel accuracy refers to the proportion of correctly segmented pixels over the total number of pixels. Precision measures the accuracy of the classifier in correctly identifying relevant pixels, such as those belonging to the object of interest in the image. A high precision indicates that positive predictions are likely to be correct, reducing false alarms or incorrect detections. Sensitivity measures the model's ability to correctly identify positive pixels out of all pixels belonging to the target class. In this study, the U-Net segmentation model achieved 98.23%-pixel accuracy, 98.40% precision, and 98.25% sensitivity. In object-level metrics, the objects being evaluated are blood cells. IoU measures the ratio of the intersection area between the predicted object and the ground truth object to the union area of both regions. The Dice coefficient is another measure of overlap between the predicted and ground truth regions, calculated as twice the intersection of the predicted and ground truth regions divided by the sum of their areas. The model achieved 95.97% IoU and 97.92% Dice coefficient. These metrics demonstrate the model's effectiveness in accurately delineating blood cell objects in the images. Table 5 displays the performance metric scores for the segmentation model.

Table 5. Performance Metrics for U-Net Model, Including Pixel-Level and Object-Level Evaluations

| Evaluation Metrics | Scores (%) |
|---|---|
| Accuracy | 98.23 |
| Precision | 98.40 |
| Sensitivity | 98.25 |
| IOU | 95.97 |
| Dice Coefficient | 97.92 |

Our CNN model was trained specifically on single blood cells extracted from comprehensive microscopic blood cell images utilizing the watershed algorithm. The dataset comprised a total of 10,614 images distributed across nine distinct classes. Figures 9 and 10 present an overview of the model's performance throughout the training process across all five-fold data. The training and validation accuracy versus epoch curve serve as an indicator of the model's learning efficacy, with ascending accuracy suggesting iterative improvement. Meanwhile, the loss versus epoch curve monitors the model's error minimization trajectory during training, with a decreasing trend indicative of effective learning while mitigating overfitting risks. Precision versus epoch and recall versus epoch curves delineate the model's proficiency in accurately identifying positive samples and capturing all relevant positive instances, respectively, both pivotal in domain-specific performance assessments. Lastly, the F1 score versus epoch curve amalgamates precision and recall metrics, furnishing a balanced assessment of model performance. Continuous monitoring of these metrics facilitates a nuanced understanding of model behavior, thereby guiding strategic refinements to architecture, hyperparameters, and training protocols to optimize overall performance and generalization capabilities.

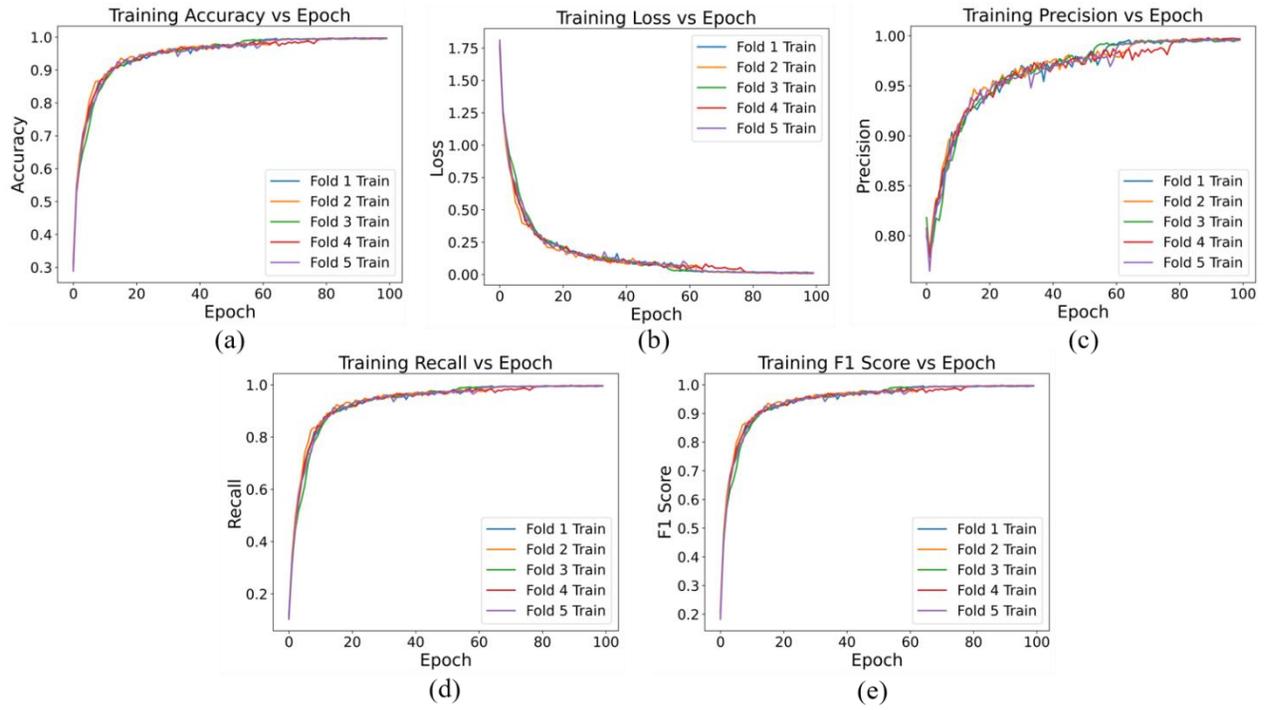

Fig 9. Training profile of the proposed LWCNN model, (a) Accuracy vs epoch, (b) Loss vs epoch, (c) Precision vs epoch, (d) Recall vs epoch, (e) F1 Score vs epoch.

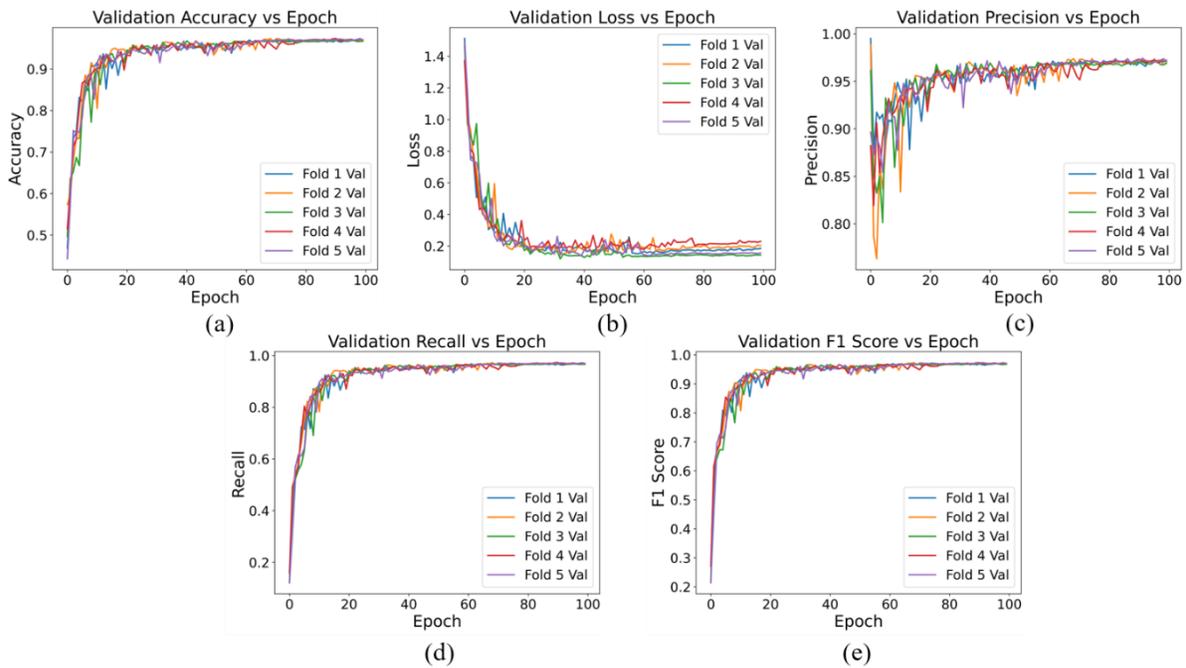

Fig 10. Validation profile of the LWCNN model, (a) Accuracy vs epoch, (b) Loss vs epoch, (c) Precision vs epoch, (d) Recall vs epoch, (e) F1 Score vs epoch.

In a classifiers model, ROC curves offer valuable insights into the performance of a CNN model during training. These curves illustrate the model's ability to discriminate between different types of blood cells, showcasing the trade-off between sensitivity (the ability to correctly identify positive instances) and specificity (the ability to correctly identify negative instances) at various threshold settings. Analyzing the ROC curve allows us to gauge how well the CNN model distinguishes between different blood cell types, with curves that hug the upper-left corner indicating superior performance. This understanding guides the selection of decision thresholds tailored to the specific requirements of blood cell classification tasks, such as prioritizing sensitivity to minimize misclassification of diseased cells or prioritizing specificity to minimize misclassification of healthy cells. Moreover, the area under the ROC curve (AUC) provides a comprehensive measure of the CNN model's overall discriminatory ability across all threshold settings, facilitating informed decision-making in blood cell classification applications. Fig. 11 displays the ROC curves for all five-fold data along with the corresponding AUC values for each class, as well as the micro-average AUC and macro-average AUC.

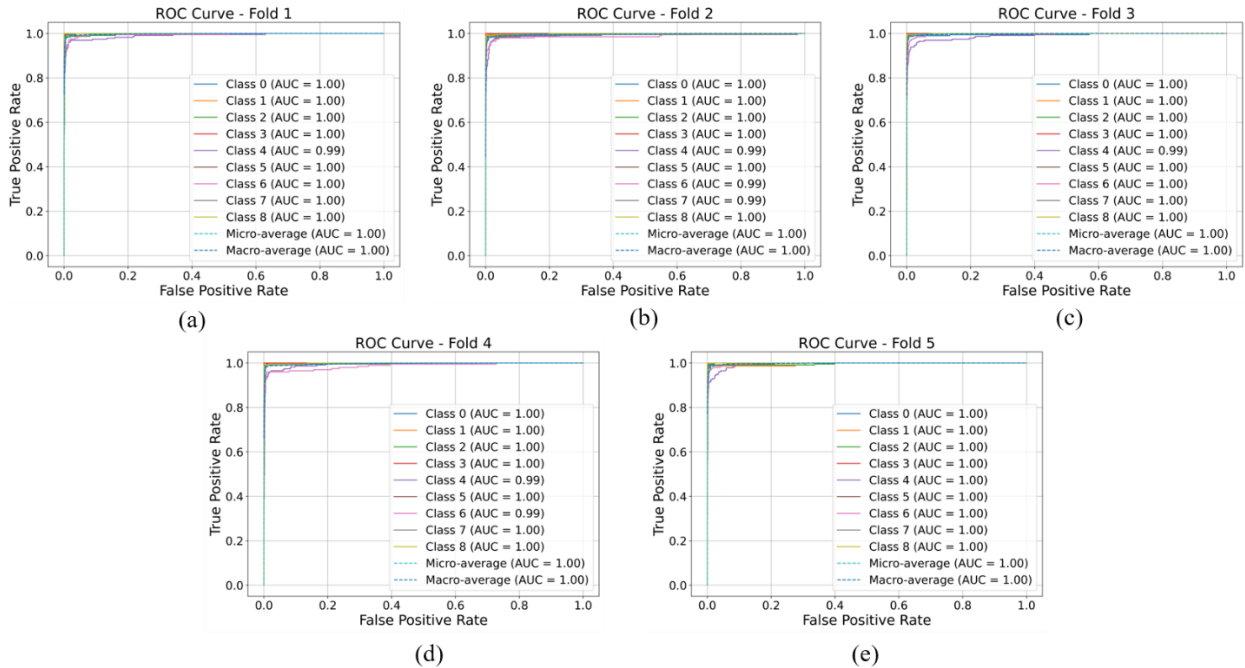

Fig 11. Receiver operating characteristic curve for (a) Fold-1, (b) Fold-2, (c) Fold-3, (d) Fold-4 and (e) Fold-5.

The performance of a CNN classifier can be evaluated using several evaluation metrics, including accuracy, precision, recall, and F1 score. Accuracy measures the proportion of correctly classified instances out of the total number of instances. However, in medical image analysis, precision and recall are particularly crucial metrics. Precision represents the ratio of correctly identified positive cases to all cases classified as positive, emphasizing the model's ability to avoid false positives. Recall, on the other hand, measures the ratio of correctly identified positive cases to all actual positive cases, highlighting the model's capability to capture all relevant instances of a particular class, thus minimizing false negatives. For instance, in medical image analysis, where the consequences of missing a diagnosis (false negative) or making an incorrect diagnosis (false positive) can be severe, the F1 score provides a comprehensive assessment of the classifier's performance. A high F1 score indicates that the classifier has achieved both high precision and high recall, striking an optimal balance between minimizing false positives and false negatives. Table 6 presents the performance metrics of our proposed CNN model across all folds and their respective averages. It is noteworthy that the model achieved an average accuracy of 97.10%, precision of 97.19%, recall of 97.01%, and F1 score of 97.10%.

Table 6. Performance metrics for the proposed CNN classifier model for 5-fold cross validation.

| Fold Number | Accuracy (%) | Precision (%) | Recall (%) | F1 Score (%) |
| --- | --- | --- | --- | --- |
| Fold-1 | 97.07 | 97.11 | 96.92 | 97.01 |
| Fold-2 | 97.41 | 97.45 | 97.31 | 97.37 |
| Fold-3 | 96.72 | 96.87 | 96.72 | 96.79 |
| Fold-4 | 97.46 | 97.60 | 97.41 | 97.50 |
| Fold-5 | 96.82 | 96.96 | 96.72 | 96.83 |
| Average | 97.10 | 97.19 | 97.01 | 97.10 |

The confusion matrix serves as a valuable tool for assessing the model's performance by providing a detailed breakdown of predicted versus actual classifications for each class. Each row in the confusion matrix corresponds to the actual class, while each column represents the predicted class. For example, if we consider the class "Basophil" as an actual class and "Eosinophil" as a predicted class, the corresponding cell in the confusion matrix would indicate the number of instances where a Basophil was incorrectly classified as an Eosinophil. By examining the diagonal elements of the confusion matrix, which represent correct classifications, and off-diagonal elements, which represent misclassifications. In Fig. 12, the confusion matrices for all folds of data are presented.

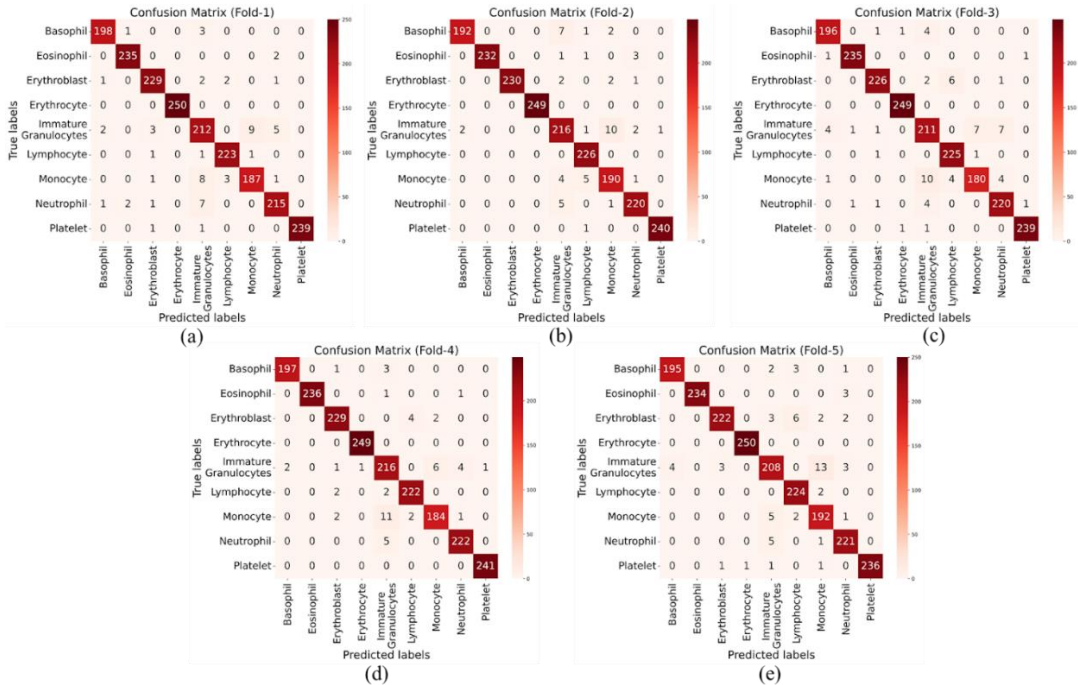

Fig 12. Confusion matrix of the LWCNN model on test set for (a) Fold-1, (b) Fold-2, (c) Fold-3, (d) Fold-4, (e) Fold-5.

A compact portion of the test dataset is depicted in Fig. 13, detailing the actual class, predicted class, and the confidence score provided by the model. Notably, all cells are correctly identified with over 99% confidence by the model.

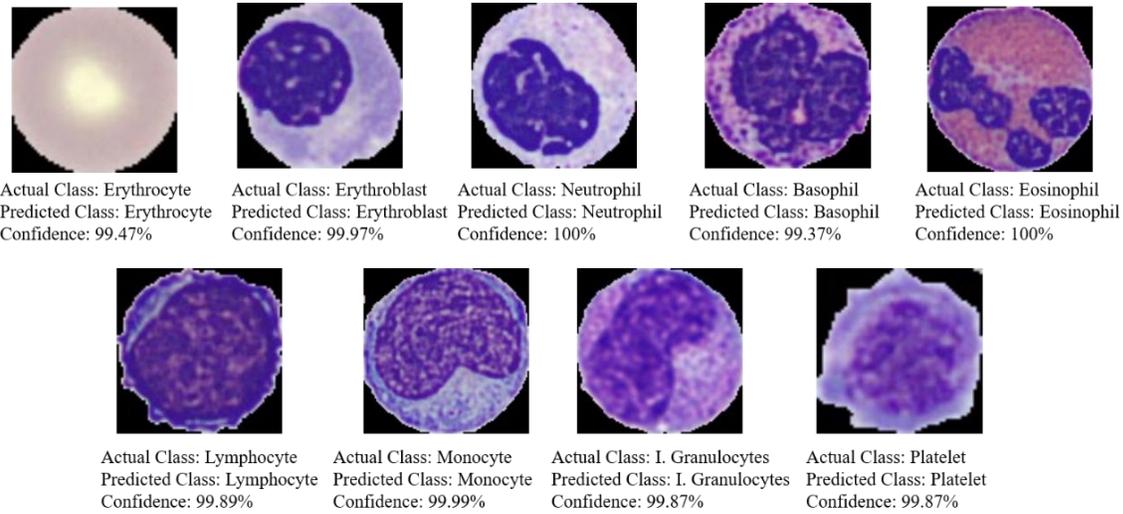

Fig 13. A representative subset of the test dataset, showcasing the confidence scores assigned by the classifier model for classification purposes.

Blood cell classification and counting from microscopic blood smear images are crucial for medical experts to make decisions aimed at preventing various blood-related illnesses. Hence, a computer-aided system is required for blood cell classification and counting to automate and ensure accuracy throughout the entire process. To achieve this objective, we proposed a comprehensive framework comprising preprocessing of microscopic peripheral blood cell images, segmentation, ROI extraction, and classification model deployment. In addition to resizing and rescaling, we implemented an image enhancement technique called CLAHE to enhance the contrast of our dataset. Given that the images in the dataset were captured under varying lighting conditions, CLAHE was particularly useful in standardizing the contrast across the dataset, thereby ensuring more consistent and reliable feature extraction and analysis. We employed data augmentation techniques to generalize the deep learning models and enhance their robustness. A CNN-based segmentation model called U-Net was utilized to segment the blood cells from the background. This step aids in isolating the blood cells from unwanted backgrounds for ROI extraction. Pixel-level metrics such as accuracy, precision, and sensitivity, along with object-level metrics like IOU and Dice Coefficient, were considered to measure the performance of the U-Net model. This model achieved 98.23% accuracy, 98.40% precision, 98.25% sensitivity, 95.97% IOU, and 97.92% Dice coefficient. For extracting single cells from the segmented blood cell images, we employed the watershed algorithm. Overlapped blood cells are common occurrences in microscopic blood smear images, and this algorithm also aids in extracting these overlapped cells. A custom-CNN architecture is proposed for classifying the single cells for counting. The nine type of blood cells are classified named Erythrocyte, Erythroblast, Neutrophil, Basophil, Eosinophil, Lymphocyte, Monocyte, Immature Granulocytes and Platelet. A 5-fold cross validation data splitting technique is applied to reduce the overfitting problems and enhance the model's robustness across the whole dataset. Several evaluation metrics like accuracy, precision, recall and f1 score are considered for measuring the model's performance. This model attained 97.10% accuracy, 97.19% precision, 97.01% recall and 97.10% f1 score. Table 7 has concluded a comparative study with the state of the art.

Table 7. Comparative analysis with state-of-the-art models in blood cell classification.

| Ref. | Classifier | Dataset | Accuracy |
|---|---|---|---|
| [90] | AlexNet, VGG Net 13, 11, ResNet 18, 34, 50, SqueezeNet 10, 11, DenseNet 121, 161. | BCCD dataset (four classes) | 100% (DenseNet 161) |
| [92] | DT, RF, SVM, k-NN, CNN, VGG16, ResNet15 | IEEE data port (3539images) Five classes | 97% (RF) |
| [91] | CNN, ELM, AlexNet, GoogleNet, VGG-16, and ResNet | BCCD dataset (four classes) | 96.03% |
| [89] | CNN | Kaggle | 98.55% |

| [88] | Optimized CNN model | IEEE data port (3539images) (Five classes) | 99.86% |
|---|---|---|---|
| [86] | CNN based MicrocellNet | Mendeley dataset (Eight classes) | 97.65% |
| [87] | Hybrid method combination of the Inception module, pyramid pooling module (PPM), and depth wise squeeze-and-excitation block | 1. Mendeley dataset (Eight classes) 2. IEEE data port (3539images) (Five classes) 3. BCCD dataset (four classes) | 1. 99.72% 2. 99.22%, 3. 99.96% |
| **Our Proposed** | **BloodCell-Net Approach (LWCNN)** | Preprocessed Blood cells (**9 classes, included erythrocytes**) Source: Mendeley dataset | **97.10%** |

## 8. Conclusion

Classifying and counting blood cells are fundamental laboratory tests essential for hematologists to make informed medical decisions. However, the manual process of classifying and counting blood cells is time-consuming, prone to errors, and labor-intensive. To address this challenge, a deep learning-based blood cell classification BloodCell-Net framework has been proposed. Several preprocessing steps, including image resizing, rescaling, histogram equalization, and data augmentation, are employed to enhance the performance of the segmentation model. A deep learning-based image segmentation technique called U-Net is utilized for segmenting the microscopic blood smear images to drop the complex and noisy background. In this research we utilized the standard U-Net architecture. Subsequently, a watershed algorithm is deployed to extract the single blood cells including overlapping cells. A custom-LWCNN model is proposed for classifying the single blood cells for counting. The proposed approach is the first study over the previous one that has included nine types of blood cells and classified them with the highest performance. Future research can be carried out on deployment of BloodCell-Net approach and a comprehensive clinical validation can be performed. This study will open a new dimension of the researcher in medical imaging.

**Data Availability:** The dataset will be published upon the request of the corresponding author. Email: simulhasantalukder@gmail.com

**Conflict of Interest**: There is no conflict of interest among the authors.

**Contribution of the Authors: SKM:** Implementation & Manuscript drafting; **MSHT**: Original manuscript drafting, Conceptualization, Data Acquisition, and supervising; **MA**: Data analysis and validation; **RBS**: Reviewing and mentoring; **MMST**: Result check and validation; **AAA**: Reviewing, grammar and sentence correcting and guiding.